\documentclass{article}
\usepackage[utf8]{inputenc}
\usepackage{hyperref}
\usepackage{xcolor}
\usepackage{amsmath}
\usepackage{siunitx}
\usepackage{textcomp}
\usepackage{comment}
\usepackage{booktabs}
\usepackage{caption}
\usepackage{subcaption}
\usepackage{longtable}
\usepackage{graphicx}
\newcommand{\subreddit}[1]{\texttt{r/{#1}}}
\usepackage{graphicx}
\graphicspath{{Figures/}}

\title{Drifts and Shifts: Characterizing the Evolution of Users Interests on Reddit}
\author{Carlo M. Valensise\textsuperscript{1,*}, Matteo Cinelli\textsuperscript{2}, Alessandro Galeazzi\textsuperscript{3},\\Walter Quattrociocchi\textsuperscript{4}}

\date{}

\begin{document}

\maketitle

\begin{center}
\textit{
\textsuperscript{1}Department of Physics, Politecnico di Milano,\\
\textsuperscript{2}Applico Lab, CNR-ISC,\\
\textsuperscript{3}Department of Information Engineering, University of Brescia, \\
\textsuperscript{4}Department of Environmental Sciences, Informatics and Statistics, University of Venice ``Ca' Foscari".
}\\
\vspace{0.75cm}
\textsuperscript{*}carlo.valensise@polimi.it

\end{center}

\begin{abstract}
Selective exposure is the main driver for the economy of attention when consuming online contents.
We select information adhering to our system of beliefs and ignore dissenting information. However, even personal interest is likely to play a role in determining our attention patterns.
To understand in more detail the dynamics of interest-driven choices, we address the evolution of users' interest on Reddit by means of an analysis on more than 7 million of users. 
We observe that the lifetime distribution of users on subreddits (online thematic communities) is 'short' with respect to the observation period.  
Furthermore, users tend to be active on a very limited number of subreddits with respect to the wide offer of the platform. These aspects indicate the presence of a migrating behavior of users among subreddits.
To characterize this phenomenon we propose a metric based on a geometrical encoding of the 'interest space' of the user. The movement of users across subreddits is characterized by a bursty trend, made of sudden variations. The most frequent of them take place between recreational subreddits and those more related to news and politics.
We describe this kind of activity with two behaviors called \emph{interest drift} and \emph{interest shift}. Our results suggest that selective exposure and personal interest coexist in driving content consumption.

\end{abstract}

\section{Introduction}

The attention of users online is a valuable asset whose dynamics are only partially clear. When the contents consumed by users pertain to a certain narrative, such as in the case of political news, it has been pointed out selective exposure dominates users' attention patterns ~\cite{bakshy2015exposure, cinelli2019selective}. We tend to select information adhering to our system of beliefs and to ignore dissenting information~\cite{del2016spreading,zollo2017debunking,zaccaria2019poprank,brugnoli2019recursive}. However, other factors may influence content selection, especially if we consider the huge and heterogeneous volume of recreational material present online. The consumption of such material is instead more likely to be driven by our own interest~\cite{dubin1992central, stebbins2001serious,bessi2017everyday}, i.e. by our desire to be involved in something we like.

The dynamics of our interests go even beyond ruling our leisure activities being expressions of our personality~\cite{holland1999interest, larson2002meta, mount2005higher,wu2007novelty}, and consequently of our culture~\cite{hofstede2004personality}, as well as important predictors of our choices~\cite{yin2016adapting}. In such a framework, the advent of the internet and social media strongly incremented the possibility to find new materials inducing a change in the way and in the pace at which we develop new interests~\cite{sintas2015nature}.

Such aspects let room for the following questions: do a wider amount of options induce sudden changes to our interests? Are we able to find a balance between our curiosity and a continuously growing offer of new material? Do we experience cognitive limitations in terms of the number of interest we can actively pursue?
In such a vein, we study the dynamics of human interests by analyzing how individuals interact with a variety of online thematic communities (called \emph{subreddits}) on Reddit, one of the most used online platform for social networking\footnote{www.reddit.com}.
The social news aggregation website Reddit was founded in 2005 and has grown over the years to now have over 200M unique users. 
Being more and more populated and one of the most visited domains of the web, Reddit is gaining popularity among researches interested in studying online human behavior as well as more technical aspects of online communities~\cite{medvedev2017anatomy}. For what concerns the dynamics of human interests, in~\cite{singer2014evolution} the authors analyzed 5 years of Reddit’s lifetime from both a user and submission (i.e. post) perspective, to show how community level attention evolves over time, resulting in diversification of topics and a concentration towards a few selected topics. In~\cite{tan2015all} the authors discovered a never ending migrating behavior of Reddit users across different communities.
Along this path, by fully tracking the activity of 7M+ users on 944 communities of interest over a period of 7 months, we find that users tend to interact with a very limited number of subreddits, almost regardless their activity amount. Such evidence regarding the limitation of human interests is also confirmed by an inherent heterogeneity of the users' activity, manifested through the concentration on a very limited set of sources with respect to the available offer, that cannot be reproduced by random models.
The limitations observed in the number of sudreddits (and, by extension, in the number of interests) on which an individual can be active at the same time is in line with a stream of literature that aims at testing the presence and the effects of cognitive constraints both online and offline. 
In more detail, such an aspect confirms how cognitive constraints, enhanced by space and time limitations, affect a number of aspects of our life such as the number of people~\cite{dunbar1992neocortex, goncalves2011validation}, places~\cite{alessandretti2018evidence} and information sources~\cite{de2019strategies, cinelli2019selective} we can be in contact with.

Through a thorough quantitative analysis we investigate how users move across different communities and topics during the period of observation. 
Under the assumption that the interaction with subreddits well represents the evolution of users' interests, we note two main behaviors that we call \textit{interest drift} and \textit{interest shift}.
A drift is a sudden change of subreddit within the same topic (e.g. from \texttt{/r/NFL} to \texttt{/r/FIFA}, both belonging to \texttt{Sport}); a shift is instead a sudden change from one topic to another (e.g from \texttt{Sport} to \texttt{NewsPoliticsSociety}).
Our findings suggest that users attention on social platforms is dominated by recreational purposes. 
Indeed, focusing on the overall users activity over time, we observe a certain burstiness in the way users tend to move across the sources of their interest. This behavior is characterized by sudden variations rather than smooth changes.
The \emph{drift} dynamics may result of interest in the case of engaging/polarizing topics, and could be correlated to a segregation within the opinion space. 
The \emph{shift} dynamic may result of interest in understanding the interrelation and the sequence of the different topics followed by a user, regardless the existence of an associated narrative.
In fact, we argue that the attention of users is shaped by the coexistence of at least two factors, namely personal interest and selective exposure.  

The paper is structured as follows.  First, we present the dataset with some descriptive statistics. Next, we characterize the users' engagement on subreddits, showing that there are limitations to the number of sources a user can interact with. Then, we introduce a metric to account for the migrating behavior of users, based on a geometrical encoding. Finally, we discuss the influence of automated accounts on our analysis.

\section{The dataset}
\begin{figure}[htbp]
    \centering
    \includegraphics[width=\columnwidth]{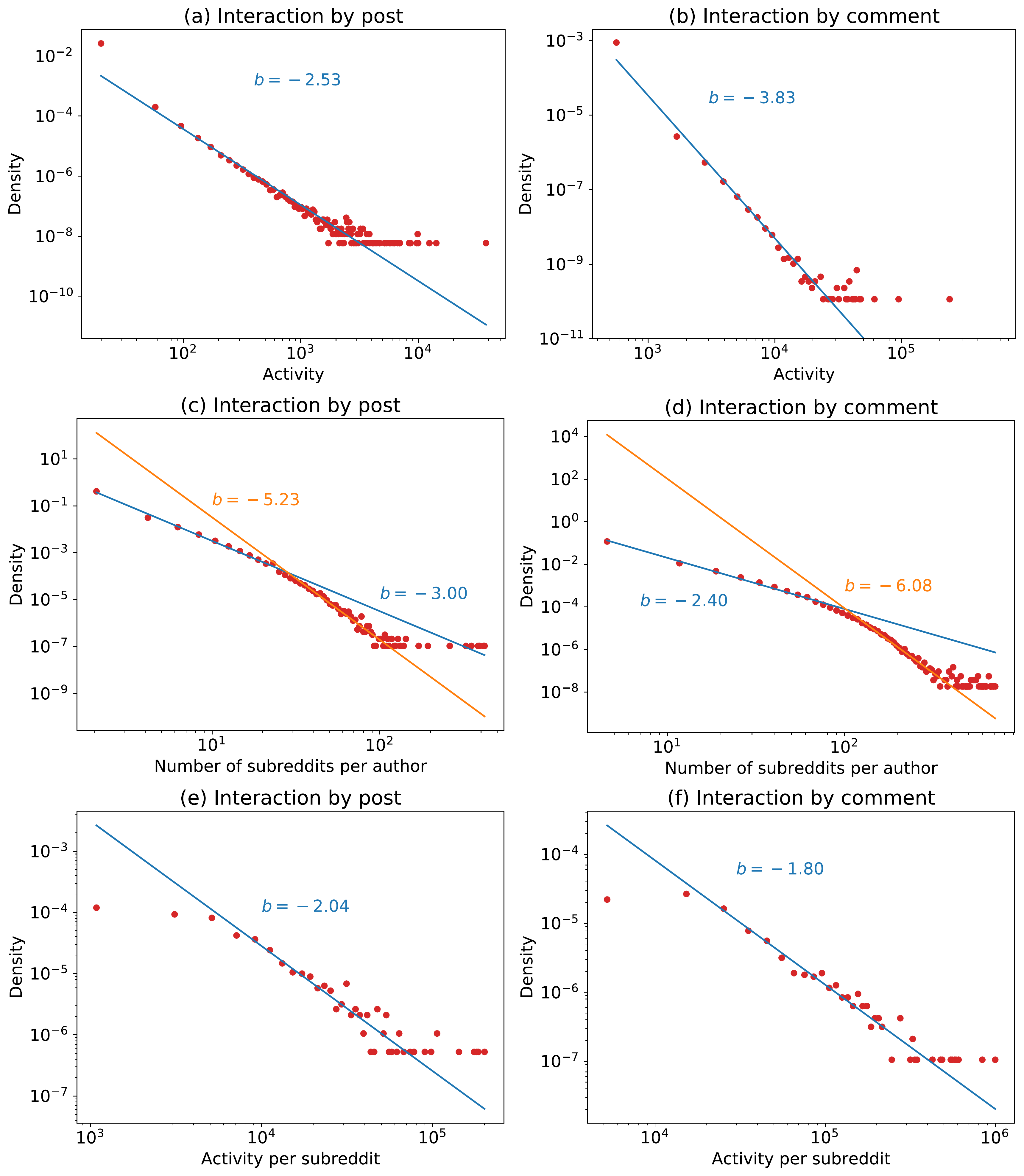}
    \caption{Authors' activities: distributions for the number of posts (a) and comments (b) per author; number of different subreddits onto which authors post (c) or commenting (d). Subreddit activities: distributions for the number of post (e) and comments (f) per subreddit; solid lines are power law fits, whose coefficient is reported in the plot as $b$.}
    \label{fig:Author_act}
\end{figure}

We exploit all the comments and posts of Reddit available from \cite{stuckInMatrix}, over 7 months, starting from the 1\textsuperscript{st} June 2018 to the 31\textsuperscript{th} December 2018. From the whole corpus, a selection is performed according to the following criteria. First, only public subreddits with a number of active users higher than 10k are considered. A further selection is performed evaluating the main topic of the subreddit. Indeed, every subreddit has a \emph{public description} that can be fed into a topic modeling algorithm~\cite{gerlach2018network} that returns the topics treated by each subreddit.
The topic modeling algorithm returns four clusters of subreddits, and five groups of topics, that are linked to the pages with different proportions. The topics are represented as groups of related words and we qualitatively identify the following semantic areas: 
\begin{itemize}
    \item visual content: images, videos and gifs;
    \item posting rules;
    \item general discussion: games, fantasy and technology;
    \item general discussion: news, politics and economics;
    \item stopwords.
\end{itemize}
We considered only those subreddits whose stronger links points towards the general discussion areas. In this way we can neglect subreddits mainly focused on visual content. 
After the selection, the corpus used for the analysis is made up of 944 subreddits, 19M+ posts, 315M+ comments, involving 7M+ users. 
The corpus has been initially evaluated from both \emph{users} and \emph{subreddits} standpoints. 
The activity (i.e. posts and comments) distributions across authors and subreddits show a power law trend, as highlighted in Figure~\ref{fig:Author_act}. As in other social networks, it appears that the majority of users produce few contents, while only a restricted number has a large production. The latter group encompasses also bots and automated accounts. Eventually, also the subreddits follow the same behaviour: apart from a few \emph{giant} pages that host a high quantity of posts and comments, the largest part shows lower levels of activity. In Figure~\ref{fig:Author_act}, for each distribution, the results of the numerical fit performed with the function $f(x) = a\,x^b$, are reported. For some of them, namely panel (c) and (d) of Figure~\ref{fig:Author_act}, a double power decay was observed. 
The shape of such distributions may indicate that, in general, users tend to be active on a limited number of subreddits.

\section{Limitations to users' interests}
\label{sec:Limitations}

\begin{figure}[htbp]
    \centering
    \includegraphics[width=\columnwidth]{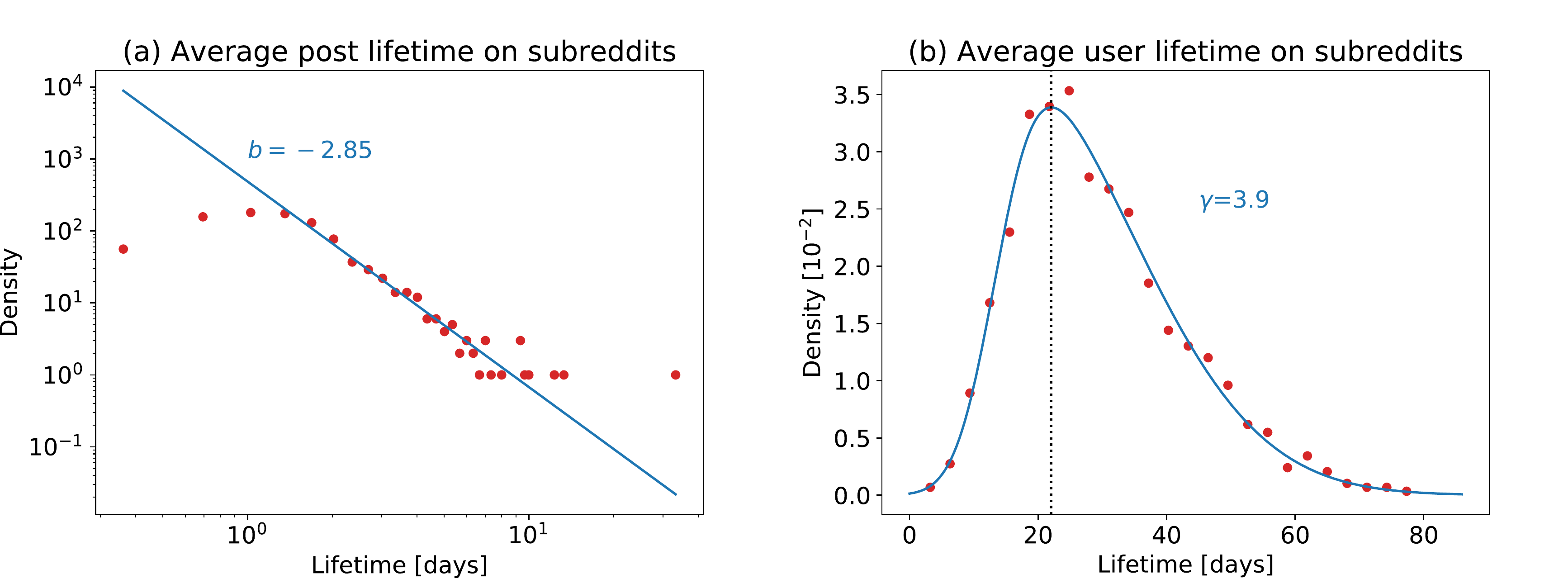}
    \caption{Lifetimes. Distribution of (a) post lifetime and (b) user lifetime per subreddit. Solid lines are, respectively, a power law fit (exponent $b$) and a skewed gaussian fit (skweness parameter $\gamma$).}
    \label{fig:Subreddit_lifetime}
\end{figure}

Among the different metadata available for each comment and for each post, the creation time, expressed in Coordinated Universal Time (UTC), with timestamps in seconds, allows us to compute the \emph{lifetime} of both posts and users on the different subreddits. 
The lifetime of a post on a subreddit is defined as the temporal distance between the first and the last comment on such a post. The lifetime of a user on a given subreddit is defined as the temporal distance between the first and the last comment on that subreddit. The lifetime of the user is defined accordingly by using his/her overall activity. Lifetime measurements are clearly limited by the period of observation.
In Figure~\ref{fig:Subreddit_lifetime} the distributions of the average post lifetime (panel (a)) and average user lifetime (panel (b)) on subreddits are reported. 
Interestingly, while the post lifetime follows a power law dependency, showing that only for a little number of posts the discussion lasts more than few days, the distribution of the user lifetime has a skewed (skewness coefficient $\gamma=3.9$)  gaussian shape, peaked around 20 days. Considering that our dataset spans 7 months, and so 210 days, this appears as a true effect and as an intriguing feature of Reddit social environment.
This suggests that users, on average, tend to move across subreddits.

\begin{figure}[htbp]
    \centering
    \includegraphics[width=\columnwidth]{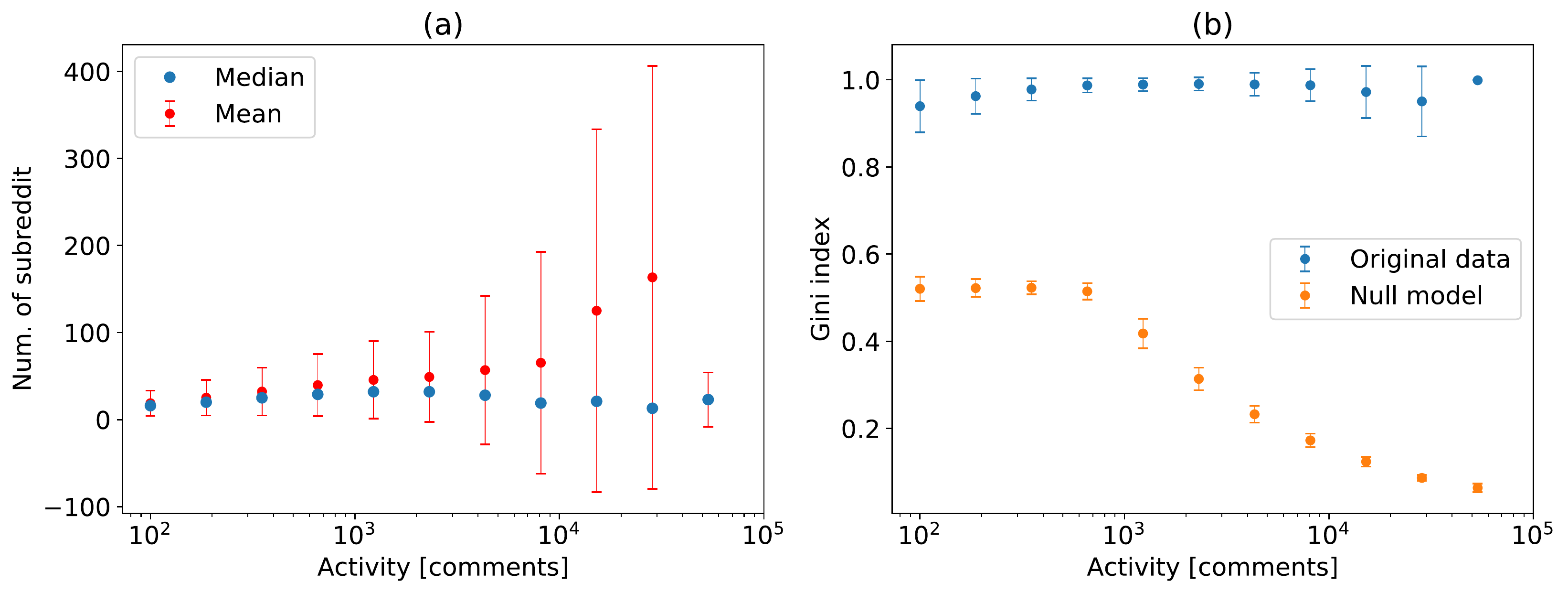}
    \caption{Users are grouped in logarithmic bins according to their activity. In panel (a), for each bin, the average number of subreddit onto which users are active is computed, together with the median. In panel (b) the average Gini index for each bin is computed, together with the average Gini index computed from the null model.}
    \label{fig:Selective_exposure}
\end{figure}

In Figure \ref{fig:Selective_exposure} (a) is reported the average number of subreddits versus the activity of each user, grouped in logarithmic bins. 
By looking at the divergence of the mean with respect to the median it turns out that the distribution of subreddits becomes more and more spread out with a growing activity. 
Therefore, with respect to the overall user activity, we note that the number of sudreddits remains quite constrained being the median approximately constant.
In panel (b) is instead reported the Gini index computed as a function of the users' activity. This quantity is a typical index used in economics to measure inequality of quantities such as income by measuring heterogeneity of their distribution~\cite{gini1921measurement}.
In this context the Gini index allows us to quantify the distribution of users' interest among different subreddits. For each user $u$, we compute a vector $\mathbf v^{(u)}$, where each element $v^{(u)}_i$ is the number of comments of user $u$ on subreddit $i$. The total number of comments for user $u$ is defined as $I_u=\sum_{i=1}^N v^{(u)}_i$. Given the total number of subreddits $N$, the Gini index for each user is computed as
\begin{equation}
g_u = \frac 1{2N} \frac{\sum_{i=1}^N\sum_{i=1}^N \lvert v^{(u)}_i-v^{(u)}_j \rvert}{I_u}
\label{eqn:gini}
\end{equation}
Values of $g\sim 1$ signal that the user concentrates his/her activity on a few subreddits only, while values of $g\simeq 0$ indicate that the activity is homogeneously split across different communities.
Being the average number of pages per user rather little with respect to the whole number of subreddit considered in this study, the activity vector $\mathbf v^{(u)}$ tends to be very sparse. To counterbalance for this effect, we consider a normalization of the Gini index for sparse data through the following:
\begin{equation}
\hat {g_u} = \frac{g_u-g^*}{1-g*}\qquad \text{where}\qquad
g^* = 
\begin{cases}
\frac {N-I_u}{I_u}\quad \text{if} \quad I_u<N\\
0 \qquad \text{otherwise}
\end{cases}
\label{eqn:gini_norm}
\end{equation}
We note an approximately constant tendency of users to be concentrated on few subreddits even with growing activity. Such a trend entails that despite their amount of activity on the social media, user tend by nature to focus on low number of subreddits, thus confirming what suggested by the panel (a) of Figure~\ref{fig:Selective_exposure}. 
We compare such a tendency with the outcomes of a null model that generates per each author a vector $\tilde {\mathbf v}^{(u)}$ that contains a number of interaction equal to $I_u$, randomly distributed across up to $I_u$ subreddits.
In other words, the null model keeps the activity of the user while reshuffling the target subreddits. After such a randomization the trend of the Gini coefficient cannot be reproduced, meaning that the heterogeneity of real data is the outcome of a specific human behaviour. In other words, the strong concentration of users seems to be the result of deliberate actions rather than random choices. Obviously, such a concentration does not mean that users do not tend to explore the media but rather that such exploration could results either in a rapid drop out (spiky dynamic) or in a strong increase of activity in a new subreddit.

\section{Evolution of Users' Interests}

The results from Section~\ref{sec:Limitations} highlight a relatively localized behaviour of users that tend to concentrate their activity on a low number of subreddits while displaying a relatively short average lifetime. The coexistence of these two conditions suggests the presence of a rather complex behaviour ruling the dynamics of users' interests that appears to go even beyond a 'simple' attitude towards wandering~\cite{tan2015all}. For instance, a user may migrate from one sport-related subreddit to another in occurrence of the end of the season thus migrating within the same domain. Another user could follow several subreddits belonging to different domains at the same time. Another one may display sudden changes of interest in terms of both subreddits and topics and so on.
For this reason, we aim at characterizing in detail the dynamics of users' interests in terms of both subreddit migration and change of topic by looking at the evolution of users' interests over time.

\begin{figure}[htbp]
    \centering
    \includegraphics[width=.7\columnwidth]{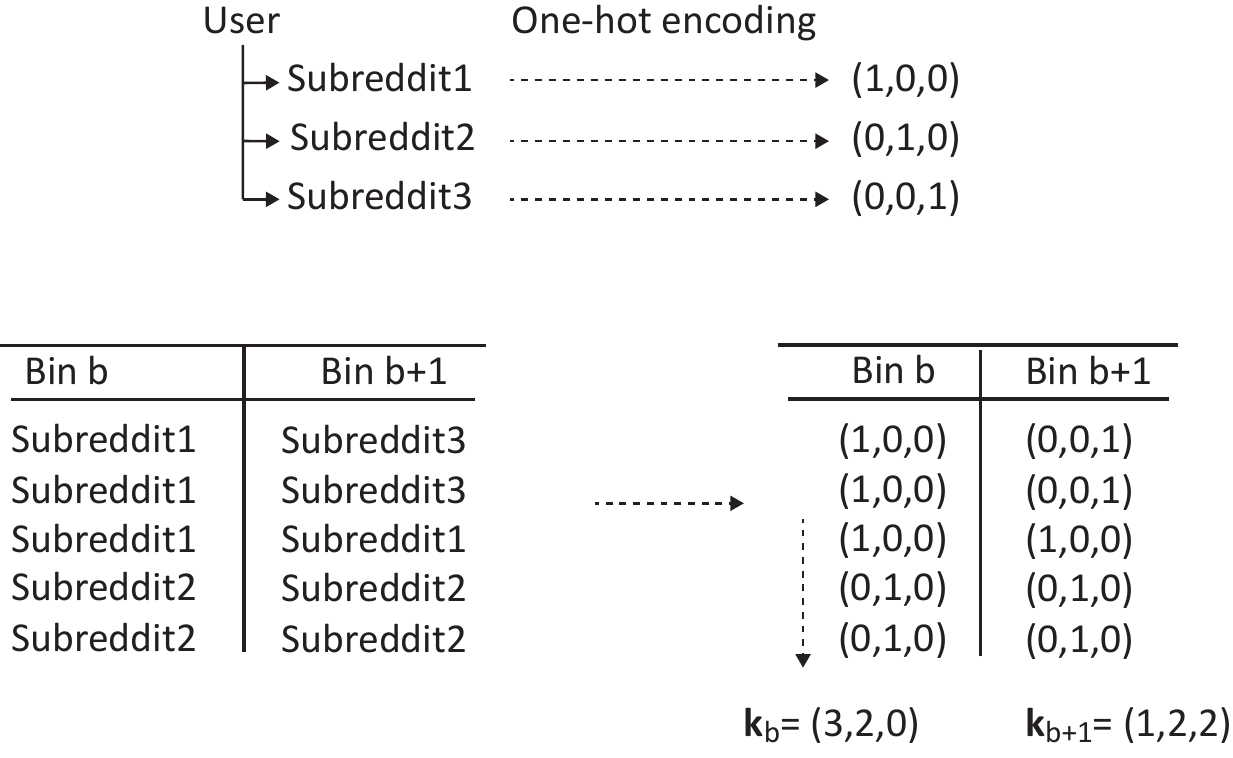}
    \caption{Activity encoding. Top part: each subreddit is transformed into a vector with one element set to 1 (active) and all the others set to 0 (one-hot encoding). Bottom part: the activity of the user is divided in bins. The corresponding vectors are summed to obtain a representation of the user activity in the bin $\mathbf k_b$. }
    \label{fig:sketch}
\end{figure}

\begin{figure}[htbp]
\centering
\includegraphics[width =\textwidth]{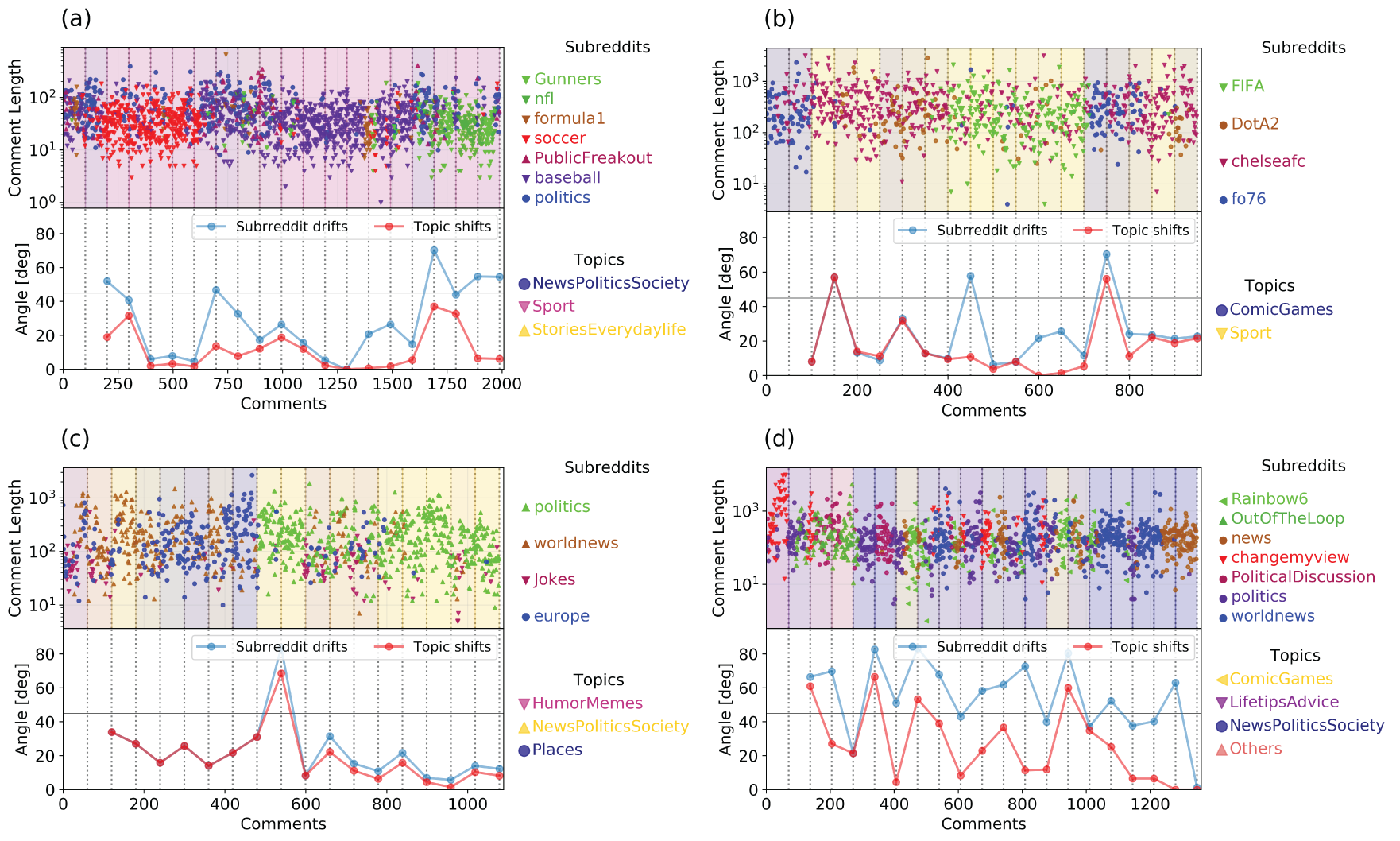}
\caption{Exemplary cases of user behavior. Along the $x$ axis the comments are reported sequentially. Top panel. Each comment is reported with its length; the color of each dot is related to the subreddit, while the marker to the topic. The background color of each bin is given by the average of topic colors. Bottom panel. The angles between consecutive vectors are reported for subreddits (blue) and topics (red).}
\label{fig:panel_users}
\end{figure}

\begin{figure}[htbp]
    \centering
    \includegraphics[width = .7\textwidth]{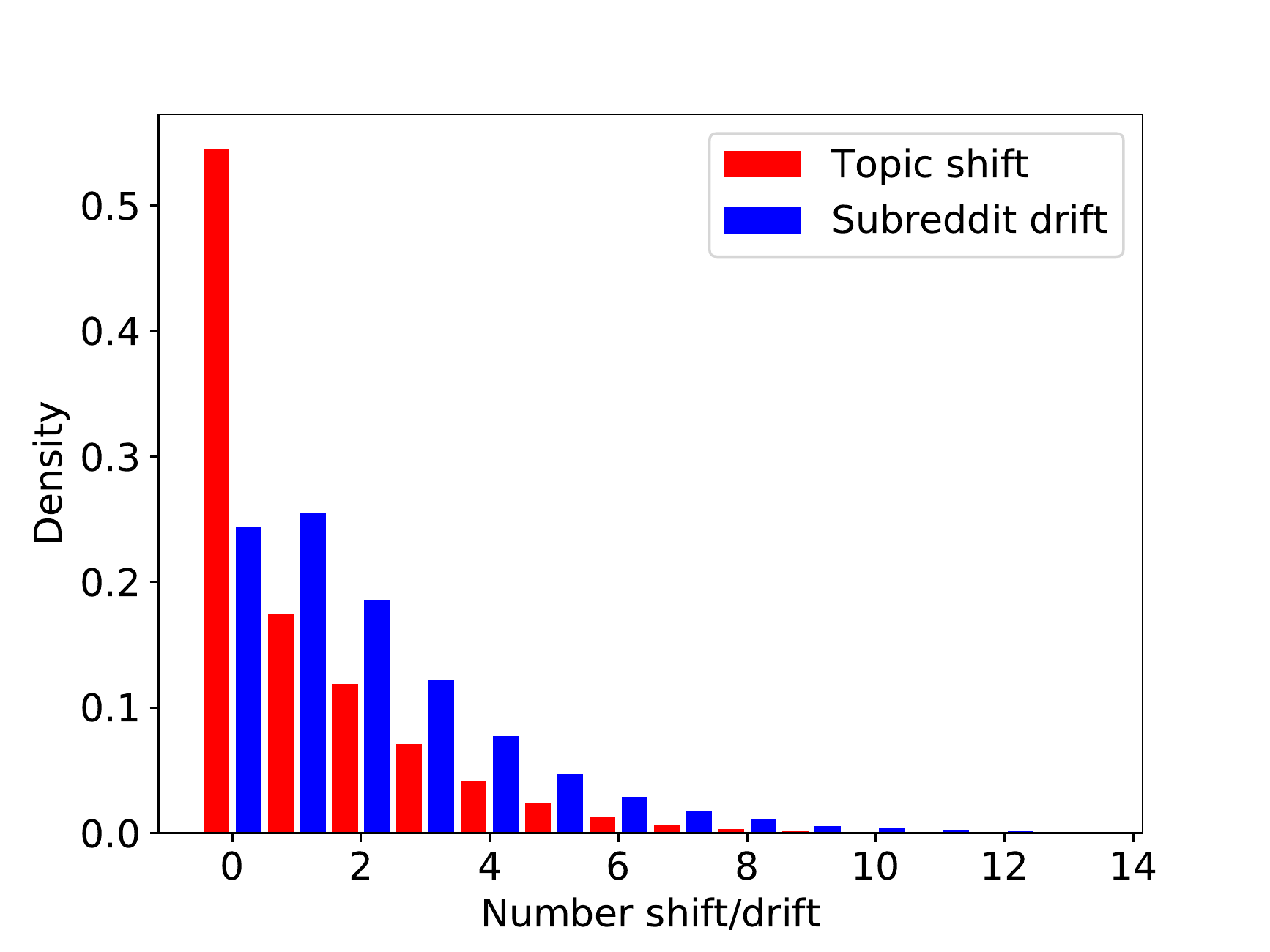}
    \caption{Topic shifts (red) and subreddit drift (blue) distributions.}
    \label{fig:shift_drift_hist}
\end{figure}

Among the 7M+ users of our dataset we follow the activity of the most active only. For each subreddit, we select the users with more than 60 comments, for a total of more than 86k users. Next, we classify by-hand all the 944 subreddits of our dataset in 15 categories related to different topics as reported in Table~\ref{tab:table_class}.
Such a classification is necessary to expand the result of the topic modeling algorithm used to preprocess the data, providing a higher topical specificity. 
To quantitatively characterize the variations over time of user interest, we temporally order all comments produced by a user, and bin them into groups of approximately equal size. Such a procedure allows us to obtain better bin-wise measurements, with respect to a grouping performed on actual daily basis, that may suffer from lags of activity thus producing empty or oversampled bins. For each user the followed subreddits are one-hot encoded into a vector of length equal to the overall number of subreddit accounted in his/her activity. Users' activity in each bin is represented by a vector $\mathbf k_b$ that is the sum of all the subreddit vectors belonging to the bin $b$. (see Figure~\ref{fig:sketch}).
A directional interpretation of these vectors allows us to map and reconstruct the way users tend to vary their interests in time. For each pair of successive vectors, the cosine similarity is computed as
\begin{equation}
    c = \frac{\mathbf k_b\cdot \mathbf k_{b+1}}{\lvert\lvert  \mathbf k_b\rvert\rvert_2\, \lvert\lvert \mathbf k_{b+1}\rvert\rvert_2}
    \label{eqn:cossim}
\end{equation}
and then the relative angle $\alpha$ is retrieved 
\[
\alpha = \arccos (c)\,.
\]
For each user we finally get a sequence of angles $\alpha_i$ with $i=1\dots B-1$ that allows us to follow and count interest variations. If a user doesn't change too much its activity between two successive bins, then the resulting angle will be close to $\ang 0$. Higher values of angle will occur whether a users significantly changes his/her commenting attitude or interest. 
This measurement is performed both with respect to subreddits and topics.
We identify two behaviours in terms of evolutions of interests that we call \emph{drift} and \emph{shift}. A drift is a sudden change of subreddit within the same topic (e.g. from \subreddit{NFL} to \subreddit{FIFA}, both belonging to `sport’); a shift is instead a sudden change from one topic to another (e.g. from `sport’ to `politics’).
A drift or a shift take place whether the angle between consecutive tuples becomes greater than $\ang{45}$.
Exemplary cases of users behavior are reported in Figure \ref{fig:panel_users}.
The top panel contains all the comments, reported in terms of their temporal ordering on the x axis and their length (i.e. number of characters) on the y axis. In this context, the comment length is reported just for visualization purposes. As reported below it will instead be used to assess the amount of automated users in our sample.
The color of each point corresponds to a subreddit, while the marker shape corresponds to the subreddit topic, to which also a color, different from the one used for the subreddit, is assigned. In fact, the background color of each bin corresponds to the average topic color of that bin. In the bottom panels, the angles between consecutive subreddits (blue) and topics (red) tuples are reported. Points above $\ang{45}$ trigger drifts or shifts.
In Figure \ref{fig:shift_drift_hist} are reported the distributions of the number of drifts and shifts. 
\begin{figure}
    \centering
    \includegraphics[width =\columnwidth]{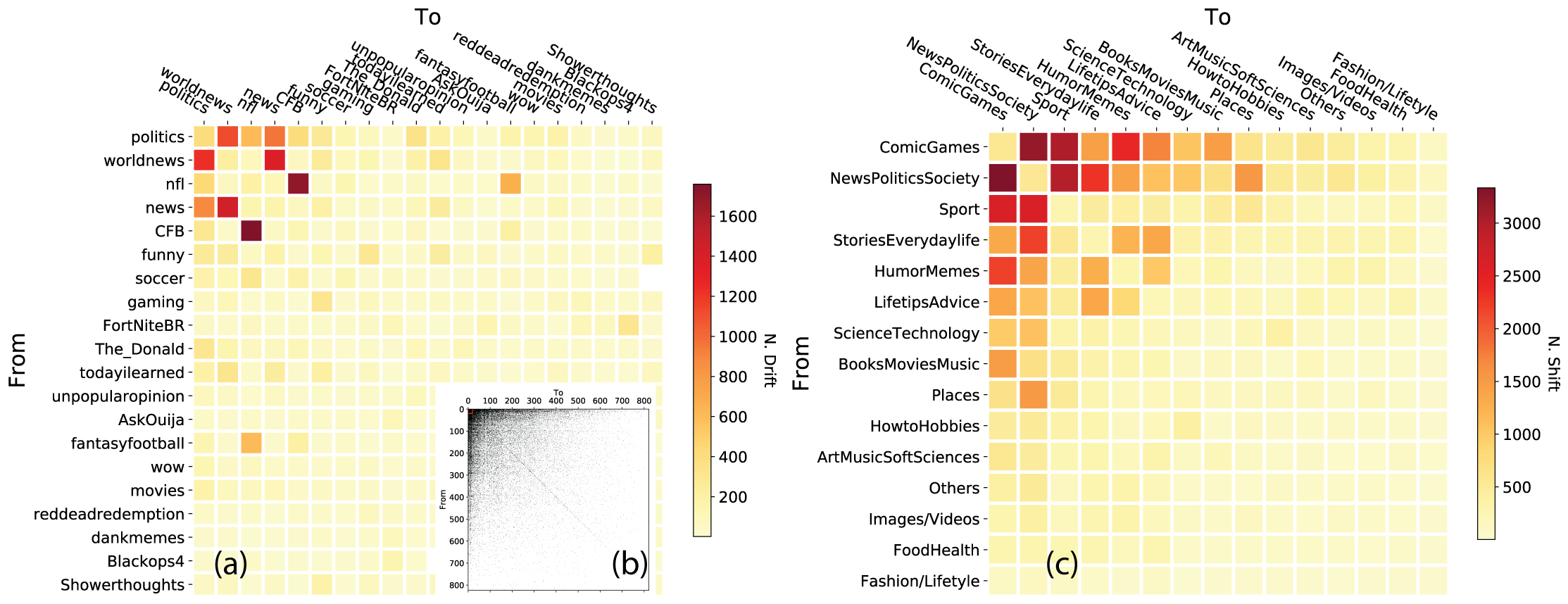}
    \caption{(a) Detail of interest drift colormap considering subreddit pairs involved in the highest number of drifts. The entire interest drift map is reported in the inset (b) as sparsity plot. (c) Colormap of interest shifts. Darkest colors correspond to a higher number of drifts/shifts. The angular measurement of topic variations allows shifts between the same class, e.g. to to an activity drop in less relevant topics. This explains why values in the diagonal may be not zero.}
    \label{fig:colormap}
\end{figure}

We note that users are more prone to a drifting behavior that entails a sudden change of subreddit or the return to an old one. Such a behavior behind the dynamic of drifts may be explained by the real discover of a new interest as well as the seasonality of certain interests determined, for instance, by sport seasons.
While drifts are relatively common, a change of topic, i.e. a shift, is more unlikely but still not impossible. Indeed 50\% of the users display at least one shift.
To get a closer look to the drift and shift dynamic, we build the colormaps shown in Figure~\ref{fig:colormap}, that reports the number of times a drift/shift between subreddits/topics occur. To enhance the plot readability the most frequent links are clustered towards the upper left corner. As descried above, and shown in Figure~\ref{fig:panel_users}, each bin contains a mixture of subreddit and topics simultaneously. To build the colormap we selected the bin pairs that trigger a drift/shift (i.e. those whose relative angle is higher than $\ang{45}$) and extracted the two most frequent elements within the two bins, counting that link as the one leading the interest change. However, interest variations can also be triggered from other events rather that insertion/removal of a new element, such as activity drops of less relevant subreddit/topics that are nevertheless able to raise the angle above the $\ang{45}$ threshold. In this case the dominant element will remain the same between the two bins, and that's why some diagonal values are not zero. A drift/shift between the same subreddit/topic is something that can occur within the definition of our metric. 
In Figure~\ref{fig:colormap} (a) a detail of the colormap for the drifts is shown, corresponding to the more frequent links. Inset (b) shows the sparsity of the adjacency matrix for all the possible links. As expected, drifts take mainly place between subreddits belonging to the same class suggesting the tendency of users to maintain their interest in a topic while varying the sources and channels onto which express it.
Figure~\ref{fig:colormap} (c) is the colormap related to the shifts. We note that certain links are more common than others (e.g. from \texttt{ComicsGames} to \texttt{NewsPoliticsSociety} and viceversa) eliciting the topics that are more subject to sudden spikes of interest. However, the symmetric appearance observed in Figure~\ref{fig:colormap} (c) seems to confirm a sort of oscillating behavior of users with respect to their interests.
We also note how users interested in certain topics such as \texttt{FoodHealth} tend to avoid shifts, keeping constant their activity with respect to the topic they are interested in. 

\section{Detection and influence of automated accounts}    
\begin{figure}[htbp]
    \centering
    \includegraphics[width =.7\columnwidth]{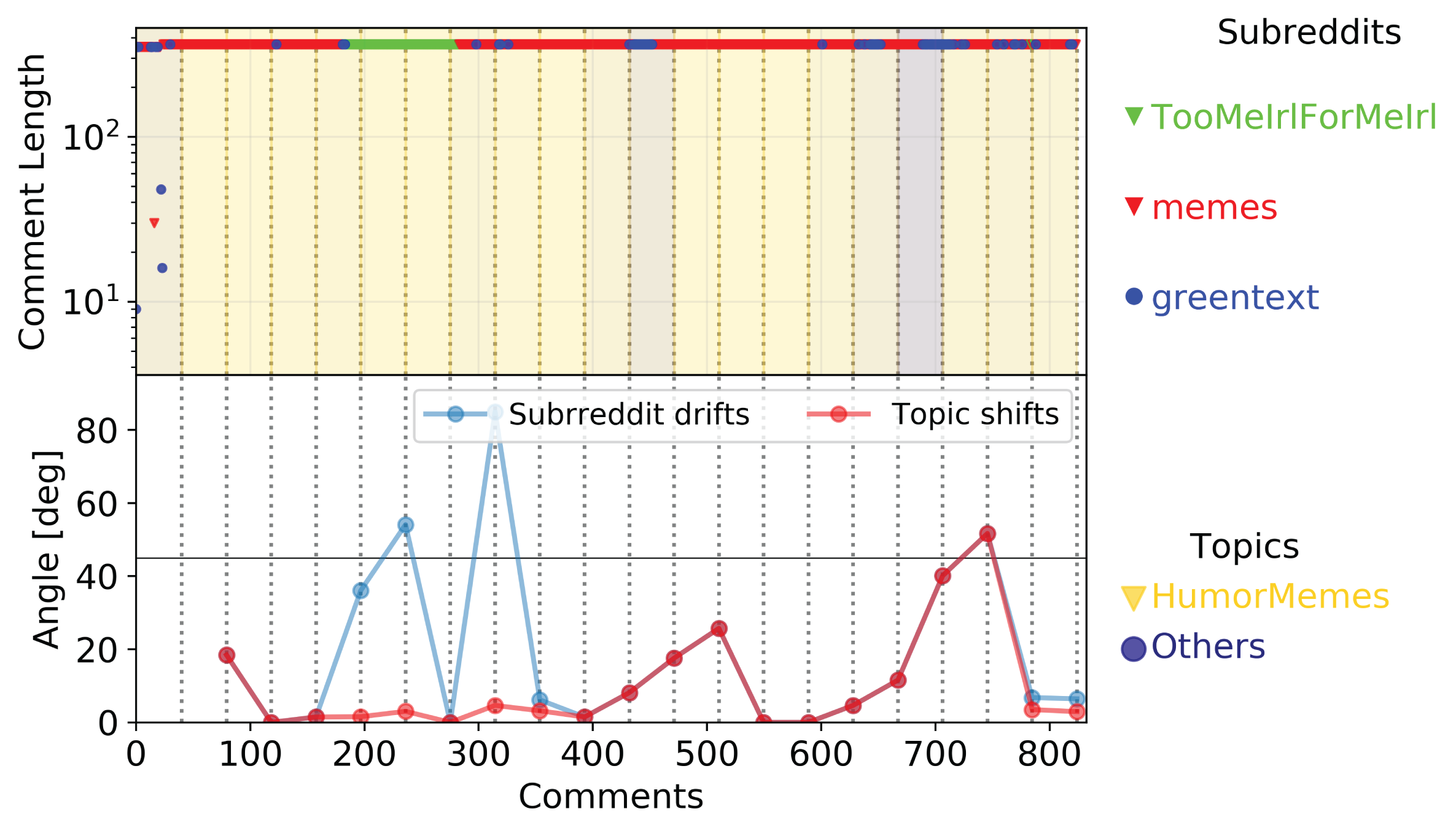}
    \caption{Activity of an automated account.}
    \label{fig:bot_example}
\end{figure}

\begin{figure}[htbp]
    \centering
    \includegraphics[width =\columnwidth]{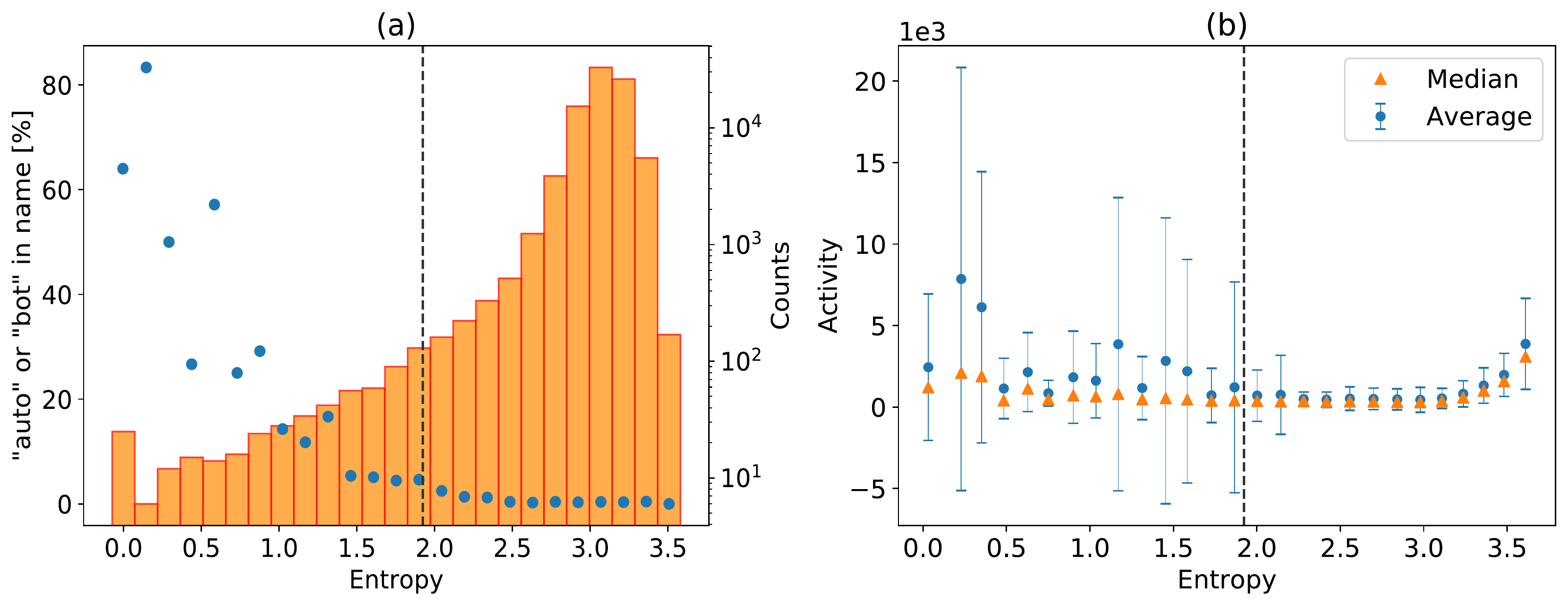}
    \caption{Bot detection. In panel (a) the distribution of entropy values is reported (right axis), together with the percentage of users with strings 'auto' or 'bot' in name per each bin (left axis, blue dots). In panel (b) the average activity is reported as a function of the entropy.}
    \label{fig:entropy_bots}
\end{figure}

As others online social networks, Reddit is populated with automated accounts. Bot detection on Reddit is beyond the scope of this paper. However, to have an intuition of the amount of automated accounts that are present in our sample, we performed the following measurements. The idea is to exploit the comment length and the total activity of an account. Accounts with a total number of comments higher than $10^4$ comments in 7 months (i.e. $\sim$48 comments per day), are likely to be automated. However, the amount of activity cannot be by itself a discriminant variable to assess whether an account is automated or not. Therefore, we considered also the variability in terms of comments length. In accordance with~\cite{cresci2017exploiting}, we assume that bots may display repetitive patterns in their comment activity resulting in a comment length distribution characterized by a low value of entropy~\cite{shannon}, as shown in Figure~\ref{fig:bot_example}. In this scenario, bots should display lower values of entropy, with respect to genuine users that should be more variable in terms of length and written number of characters.
In Figure~\ref{fig:entropy_bots}~(a) is reported the distribution of entropy values computed from the distribution of comment length per each user. We didn't account for subreddits that have programmatic exotic commenting rules such as \subreddit{AskOuija}, where users are allowed to comment with a single letter only.
This distribution shows an exponential behavior, suggesting that the largest part of users has high values of entropy (note the log scale on right axis). The blue dots, referred to the left axis, report, for each bin, the percentage of user that display the string 'bot' or 'auto' in name. These users represent the 0.5\% of the entire population, as shown by the vertical black line, corresponding to the 0.5-th percentile. Complementary to this information, panel (b) shows the activity of users as a function of the entropy. Even in this case the 0.5-th percentile is reported. Interestingly large deviations from the mean of the activity occur mainly below this threshold thus reinforcing the hypothesis that automated account show limited variability in terms of comments (i.e. low entropy). From this brief analysis we can conclude that the largest part of users in our sample are genuine accounts, and bots or other kind of automated account are in an irrelevant, but detectable percentage.

\section{Conclusions}
In this paper we quantitatively described the variation of users' interest on Reddit. Due to cognitive and physical constraints users can follow a limited number of information sources simultaneously. This aspect is highlighted by the finite lifetime of users' on subreddits, entailing that after a certain amount of time the interest over a particular subreddit drops. Moreover, the Gini index shows that, despite the large offer of thematic communities, users concentrate their activity on a very limited number of them. To get a closer look to the dynamic of migration between subreddits, and look at it also in the light of interests variation, we proposed a geometrical encoding of users' activity defining  \emph{drifts} and \emph{shifts}, that represent abrupt variations of activity with respect to subreddits and topics respectively. 
The dynamics related to users interests and in particular the study of drifts and shifts may result important for multiple reasons. In more detail, the dynamic of drifts may be especially important when related to polarizing topics such as politics. 
Indeed, a certain user whose opinion becomes more and more extreme over time could display several consecutive drifts towards subreddits with a stronger political orientation. At the same time the study of interest shifts may result important both for the detection of relevant topics in a given moment but also for other issues related to human behavior. For instance, enrolling and focusing the activity on subreddit of emotional support may signal of a moment of emotional weakness. 
The strong shifts concentration between a few topics pairs suggest a severe separation among different topics.
More interestingly, the presence of shifts confirms that users consume material pertaining to a wide variety of topics that could be linked or not to a certain narrative. In the former case, as confirmed by previous studies, the consumption could be guided by selective exposure while in the latter case it could be guided by interest only. Therefore, the coexistence of selective exposure driven and interest driven consumption of online content seems to be a more reasonable explanation for the patterns of users' attention.
Future work should be devoted to investigate in more detail the relationship between selective exposure and personal interest, introducing further aspects such as users' opinion on controversial and debated topics.

\bibliographystyle{unsrt}

\begin{thebibliography}{10}

\bibitem{bakshy2015exposure}
Eytan Bakshy, Solomon Messing, and Lada~A Adamic.
\newblock Exposure to ideologically diverse news and opinion on facebook.
\newblock {\em Science}, 348(6239):1130--1132, 2015.

\bibitem{cinelli2019selective}
Matteo Cinelli, Emanuele Brugnoli, Ana~Lucia Schmidt, Fabiana Zollo, Walter
  Quattrociocchi, and Antonio Scala.
\newblock Selective exposure shapes the facebook news diet.
\newblock {\em arXiv preprint arXiv:1903.00699}, 2019.

\bibitem{del2016spreading}
Michela Del~Vicario, Alessandro Bessi, Fabiana Zollo, Fabio Petroni, Antonio
  Scala, Guido Caldarelli, H~Eugene Stanley, and Walter Quattrociocchi.
\newblock The spreading of misinformation online.
\newblock {\em Proceedings of the National Academy of Sciences},
  113(3):554--559, 2016.

\bibitem{zollo2017debunking}
Fabiana Zollo, Alessandro Bessi, Michela Del~Vicario, Antonio Scala, Guido
  Caldarelli, Louis Shekhtman, Shlomo Havlin, and Walter Quattrociocchi.
\newblock Debunking in a world of tribes.
\newblock {\em PloS one}, 12(7):e0181821, 2017.

\bibitem{zaccaria2019poprank}
Andrea Zaccaria, Michela Del~Vicario, Walter Quattrociocchi, Antonio Scala, and
  Luciano Pietronero.
\newblock Poprank: Ranking pages’ impact and users’ engagement on facebook.
\newblock {\em PloS one}, 14(1):e0211038, 2019.

\bibitem{brugnoli2019recursive}
Emanuele Brugnoli, Matteo Cinelli, Walter Quattrociocchi, and Antonio Scala.
\newblock Recursive patterns in online echo chambers.
\newblock {\em arXiv preprint arXiv:1908.11583}, 2019.

\bibitem{dubin1992central}
Robert Dubin.
\newblock {\em Central life interests: Creative individualism in a complex
  world}.
\newblock Transaction Publishers, 1992.

\bibitem{stebbins2001serious}
Robert~A Stebbins.
\newblock Serious leisure.
\newblock {\em Society}, 38(4):53, 2001.

\bibitem{bessi2017everyday}
Alessandro Bessi, Fabiana Zollo, Michela Del~Vicario, Antonio Scala, Fabio
  Petroni, Bruno Gon{\c{c}}calves, and Walter Quattrociocchi.
\newblock Everyday the same picture: Popularity and content diversity.
\newblock In {\em International Workshop on Complex Networks}, pages 225--236.
  Springer, 2017.

\bibitem{holland1999interest}
John~L Holland.
\newblock Why interest inventories are also personality inventories.
\newblock 1999.

\bibitem{larson2002meta}
Lisa~M Larson, Patrick~J Rottinghaus, and Fred~H Borgen.
\newblock Meta-analyses of big six interests and big five personality factors.
\newblock {\em Journal of Vocational Behavior}, 61(2):217--239, 2002.

\bibitem{mount2005higher}
Michael~K Mount, Murray~R Barrick, Steve~M Scullen, and James Rounds.
\newblock Higher-order dimensions of the big five personality traits and the
  big six vocational interest types.
\newblock {\em Personnel psychology}, 58(2):447--478, 2005.

\bibitem{wu2007novelty}
Fang Wu and Bernardo~A Huberman.
\newblock Novelty and collective attention.
\newblock {\em Proceedings of the National Academy of Sciences},
  104(45):17599--17601, 2007.

\bibitem{hofstede2004personality}
Geert Hofstede and Robert~R McCrae.
\newblock Personality and culture revisited: Linking traits and dimensions of
  culture.
\newblock {\em Cross-cultural research}, 38(1):52--88, 2004.

\bibitem{yin2016adapting}
Hongzhi Yin, Xiaofang Zhou, Bin Cui, Hao Wang, Kai Zheng, and Quoc Viet~Hung
  Nguyen.
\newblock Adapting to user interest drift for poi recommendation.
\newblock {\em IEEE Transactions on Knowledge and Data Engineering},
  28(10):2566--2581, 2016.

\bibitem{sintas2015nature}
Jordi~L{\'o}pez Sintas, Laura~Rojas de~Francisco, and Ercilila~Garc{\'\i}a
  {\'A}lvarez.
\newblock The nature of leisure revisited: An interpretation of digital
  leisure.
\newblock {\em Journal of Leisure Research}, 47(1):79--101, 2015.

\bibitem{medvedev2017anatomy}
Alexey~N Medvedev, Renaud Lambiotte, and Jean-Charles Delvenne.
\newblock The anatomy of reddit: An overview of academic research.
\newblock In {\em Dynamics on and of Complex Networks}, pages 183--204.
  Springer, 2017.

\bibitem{singer2014evolution}
Philipp Singer, Fabian Fl{\"o}ck, Clemens Meinhart, Elias Zeitfogel, and Markus
  Strohmaier.
\newblock Evolution of reddit: from the front page of the internet to a
  self-referential community?
\newblock In {\em Proceedings of the 23rd international conference on world
  wide web}, pages 517--522. ACM, 2014.

\bibitem{tan2015all}
Chenhao Tan and Lillian Lee.
\newblock All who wander: On the prevalence and characteristics of
  multi-community engagement.
\newblock In {\em Proceedings of the 24th International Conference on World
  Wide Web}, pages 1056--1066. International World Wide Web Conferences
  Steering Committee, 2015.

\bibitem{dunbar1992neocortex}
Robin~IM Dunbar.
\newblock Neocortex size as a constraint on group size in primates.
\newblock {\em Journal of human evolution}, 22(6):469--493, 1992.

\bibitem{goncalves2011validation}
Bruno Goncalves, Nicola Perra, and Alessandro Vespignani.
\newblock Validation of dunbar's number in twitter conversations.
\newblock {\em arXiv preprint arXiv:1105.5170}, 2011.

\bibitem{alessandretti2018evidence}
Laura Alessandretti, Piotr Sapiezynski, Vedran Sekara, Sune Lehmann, and Andrea
  Baronchelli.
\newblock Evidence for a conserved quantity in human mobility.
\newblock {\em Nature Human Behaviour}, 2(7):485, 2018.

\bibitem{de2019strategies}
Marco De~Nadai, Angelo Cardoso, Antonio Lima, Bruno Lepri, and Nuria Oliver.
\newblock Strategies and limitations in app usage and human mobility.
\newblock {\em Scientific reports}, 9(1):10935, 2019.

\bibitem{stuckInMatrix}
\url{https://files.pushshift.io/reddit/}.

\bibitem{gerlach2018network}
Martin Gerlach, Tiago~P Peixoto, and Eduardo~G Altmann.
\newblock A network approach to topic models.
\newblock {\em Science advances}, 4(7):eaaq1360, 2018.

\bibitem{gini1921measurement}
Corrado Gini.
\newblock Measurement of inequality of incomes.
\newblock {\em The Economic Journal}, 31(121):124--126, 1921.

\bibitem{cresci2017exploiting}
Stefano Cresci, Roberto Di~Pietro, Marinella Petrocchi, Angelo Spognardi, and
  Maurizio Tesconi.
\newblock Exploiting digital dna for the analysis of similarities in twitter
  behaviours.
\newblock In {\em 2017 IEEE International Conference on Data Science and
  Advanced Analytics (DSAA)}, pages 686--695. IEEE, 2017.

\bibitem{shannon}
Claude~Elwood Shannon.
\newblock A mathematical theory of communication.
\newblock {\em The Bell System Technical Journal}, 27(3):379--423, July 1948.

\end{thebibliography}

\appendix

\section{Classification of sr}
\label{sec:classification}
    \begin{longtable}{p{3cm}p{6cm}p{3cm}}
    \toprule
    Topic & Description & Examples\\
    \midrule
 Sport & Subreddits collecting discussions about sports or supporting teams or athletes. & \subreddit{sports}, \subreddit{olympics}, \subreddit{tennis}, \subreddit{skateboarding} \\
 Food, Health & Subreddits collecting discussions about food and related issues as well as health and wellbeing related problems. & \subreddit{Health}, \subreddit{nutrition}, \subreddit{vegan}, \subreddit{Drugs}\\
 Comics, Games & Subreddis collecting discussions about comics and games both online and offline. & \subreddit{magicTCG}, \subreddit{PS4}, \subreddit{comicbooks}, \subreddit{DnD}\\
 News, Politics, Society &   Subreddits collecting discussions about news, politics (regardless the party involved) and societal issues such as migration or abortion. & \subreddit{news}, \subreddit{Conservative}, \subreddit{worldpolitics}, \subreddit{TrumpCriticisesTrump} \\
 Science, Technology & Subreddits collecting discussions about hard sciencies and technology from computers to cryptocurrencies. & \subreddit{Physics}, \subreddit{Bitcoin}, \subreddit{science}, \subreddit{techonolgy}\\
 Lifetips, Advice &  Subreddits collecting discussions about emotional issues and providing a forum where to share experiences and receive advice. & \subreddit{parenting}, \subreddit{askwomenadvice}, \subreddit{socialanxiety}, \subreddit{raisedbynarcisists} \\
 Humor, Memes & Subreddits collecting humoristic and funny objects including memes. & \subreddit{funny}, \subreddit{dankmemes}, \subreddit{madlads}\\
 Books, Movies, Music & Subreddits collecting discussions about books, movies (including tv series and shows). & \subreddit{startrek}, \subreddit{harrypotter}, \subreddit{rickandmorty}, \subreddit{TrueFilm}\\
 Images, Videos & Subreddits dedicted to collect images and videos including GIFs without a specific topic. & \subreddit{MostBeautiful}, \subreddit{tumblr}, \subreddit{EarthPorn}\\
 Fashion, Lifestyle & Subreddit collecting discussions about fashion and lifestyle. & \subreddit{sneakers}, \subreddit{beauty}, \subreddit{tattoo} \\
 Stories, Everyday life & Subreddits collecting stories from users regarding their daily experience and everyday life that are not related to existential problem or emotional issues or the request of support. & \subreddit{TalesfromCustomer}, \subreddit{unpopularopinions}, \subreddit{LucidDreaming}, \subreddit{showerthoughts}\\
 How to, Hobbies  & Subreddits dedicate to guide and support users in technical problem or for discussing their hobbies/passions. & \subreddit{everymanshouldknow}, \subreddit{knitting}, \subreddit{MechanicAdvice}, \subreddit{DIY}\\
 Art, Music, Soft Sciences & Subreddits dedicated to discussion about soft sciences including History, Philosophy etc. & \subreddit{Foodforthought}, \subreddit{Poetry}, \subreddit{classicalmusic}, \subreddit{ArtFundamentals}\\
 Places & Subreddits dedicated to collect picture and videos of specific places/location. & \subreddit{london}, \subreddit{philippines}, \subreddit{vancouver}, \subreddit{australia} \\
 Others & Subreddits that cannot be classified in the previous categories. & \subreddit{nocontext}, \subreddit{Military}, \subreddit{conspiracy}, \subreddit{therewasanattempt}\\
 \bottomrule
\caption{Classification of subreddit}
\label{tab:table_class}
\end{longtable}

\end{document}